\documentstyle[prd,preprint,tighten,aps,amssymb,newlfont,epsfig]{revtex}
\begin{document}
\draft
\preprint{
\begin{tabular}{r}
   UWThPh-1999-25
\\ PRL-TH-1999
\end{tabular}
}

\title{FIELD-THEORETICAL TREATMENT OF NEUTRINO OSCILLATIONS}

\author{W. GRIMUS}

\address{Institute for Theoretical Physics, University of Vienna,
Boltzmanngasse 5,\\ A-1090 Vienna, Austria\\
E-mail: grimus@doppler.thp.univie.ac.at}

\author{S. MOHANTY AND P. STOCKINGER}

\address{Theory Group, Physical Research Laboratory,
Ahmedabad - 380 009, India\\
E-mail: mohanty@prl.ernet.in and stocki@prl.ernet.in}  

\maketitle

\begin{abstract}
We discuss the field-theoretical approach to neutrino
oscillations. This approach includes the neutrino source and
detector processes and allows to obtain the neutrino transition or survival
probabilities as cross sections derived from the Feynman diagram
of the combined source -- detection process. In this context, 
the neutrinos which are supposed to oscillate appear as
propagators of the neutrino mass eigenfields, connecting the
source and detection processes.
\end{abstract}

\vspace{1cm}

\pacs{Talk presented by W. Grimus at the
17th \textit{International Workshop on Weak Interactions and Neutrinos}
(WIN99),
Cape Town, South Africa, January 24--30, 1999.}

\section{Introduction}

Neutrino oscillations~\cite{pon} play a central role in neutrino
physics. The most important condition for this phenomenon is given by
neutrino mixing which is described by
$\nu_{L\alpha} = \sum_j U_{\alpha j} \nu_{Lj}$
with $\alpha = e, \mu, \tau, \ldots$ and $j = 1,2,3, \ldots$
labelling neutrino flavours (types) and mass eigenfields,
respectively. All neutrino oscillation experiments are evaluated with
the formula \cite{pon}
\begin{equation}
P_{\nu_\alpha\to\nu_\beta} (L/E_\nu) =
\left| \sum_j U_{\beta j} U^*_{\alpha j} 
\exp \left( -i \frac{m^2_j L}{2E_\nu} \right) \right|^2 \,,
\label{P}
\end{equation}
where $U$ denotes the unitary mixing matrix, $L$ the distance between
source and detector and $E_\nu$ the neutrino energy. The neutrino masses
$m_j$ are associated with the mass eigenfields $\nu_j$.
It has been indicated in several publications that the standard derivation
of Eq.~(\ref{P}) raises a number of conceptual questions
(see, e.g., Ref.~\cite{rich} for a clear exposition). Some of these
questions are solved by the wave packet approach~\cite{kayser}
(see also the review~\cite{zralek} where a list of references can be
found), however, the size and form of the wave packet is not determined in
this approach and remains a subject to reasonable estimates.

The idea has been put forward to include the neutrino production and
detection processes into the consideration of neutrino
oscillations.~\cite{rich} Such an approach can be realized with 
\emph{quantum mechanics} --
in which case the neutrinos with definite mass are
unobserved intermediate states between the source and detection
processes~\cite{rich} -- or with \emph{quantum field theory} where the
massive neutrinos are represented by inner lines in a big Feynman
diagram depicting the combined source -- detection
process.~\cite{GS96,GSM99} In the following we will discuss the
field-theoretical treatment. The aims and hopes of such an approach
are the following: 1. The elimination of the arbitrariness associated
with the wave packet approach, 2. the description of neutrino
oscillations by means of the particles in neutrino production and detection 
which are really manipulated in an experiment, 3. a more complete and
realistic description in order to find possible limitations of formula
(\ref{P}) in specific experimental situations. 

Considering laboratory experiments, there are two typical situations
for neutrino oscillation experiments. The first one is \emph{decay at
rest} (DAR) of the neutrino source. Its corresponding Feyman diagram
is depicted in Fig.~\ref{DAR}. The wave functions of the source and
detector particles are localized (peaked) at $\vec{x}_S$ and
$\vec{x}_D$, respectively. The other situation is \emph{decay in
flight} (DIF) of the neutrino source as represented in
Fig.~\ref{DIF} where it is assumed that a proton hits a target
localized at $\vec{x}_T$. The detector particle sits again at
$\vec{x}_D$ but the source is not localized. In both situations the
distance between source and detection is given by 
$L = |\vec{x}_D-\vec{x}_S|$. Note that in the Feynman diagrams of
Figs.~\ref{DAR} and \ref{DIF} the neutrinos with definite mass occur
as inner lines. In the spirit of our approach, neutrino oscillation
probabilities are proportional to the cross sections derived from the
amplitudes represented by these diagrams.

\section{Assumptions and the resulting amplitude}

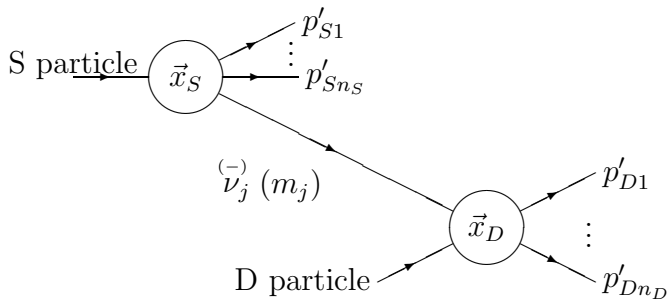
\begin{figure}[t]
\setlength{\unitlength}{5mm}
\begin{picture}(12,11)(-4,1)
\put(0,8){\vector(1,0){1}} \put(1,8){\line(1,0){1}}
\put(3,8){\circle{2}} \put(3,8){\makebox(0,0){$\vec{x}_S$}}
\put(3.8944,8.4472){\vector(2,1){1}}
\put(4.8944,8.9472){\line(2,1){1}}
\put(4,8){\vector(1,0){1}} \put(5,8){\line(1,0){1}}
\put(3.8944,7.5528){\vector(2,-1){3.1056}}
\put(7,6){\line(2,-1){3.1056}}
\put(11,4){\circle{2}} \put(11,4){\makebox(0,0){$\vec{x}_D$}}
\put(11.8944,4.4472){\vector(2,1){1}}
\put(12.8944,4.9472){\line(2,1){1}}
\put(11.8944,3.5528){\vector(2,-1){1}}
\put(12.8944,3.0528){\line(2,-1){1}}
\put(8.1056,2.5528){\vector(2,1){1}}
\put(9.1056,3.0528){\line(2,1){1}}
%
\put(0,8.1){\makebox(0,0)[b]{S particle}}
\put(6.0944,9.4472){\makebox(0,0)[l]{$p'_{S1}$}}
\put(6.2,8){\makebox(0,0)[l]{$p'_{S{n_S}}$}}
\put(5.8,8.3){\makebox(0,0)[b]{$\vdots$}}
\put(6.9,5.9){\makebox(0,0)[tr]{
$\stackrel{\scriptscriptstyle (-)}{\nu}_{\hskip-3pt j}(m_j)$
}}
\put(7.9056,2.5528){\makebox(0,0)[r]{D particle}}
\put(14.0944,5.4472){\makebox(0,0)[l]{$p'_{D1}$}}
\put(14.0944,2.5528){\makebox(0,0)[l]{$p'_{D{n_D}}$}}
\put(13.6,4){\makebox(0,0)[l]{$\vdots$}}
\end{picture}
\caption{\label{DAR} Feynman diagram for decay at rest (DAR) of
the neutrino source particle. The source (S) and detector (D) processes
are symbolized by the circles. The labels $\vec{x}_S$ and
$\vec{x}_D$ represent the coordinates where the wave functions
of the source and detector particles are peaked, respectively.
We have also indicated the $n_S$ ($n_D$) momenta of the final particles
originating from the source (detector) process and the neutrino
propagator of the neutrino field with mass $m_j$.}
\end{figure}

\begin{figure}[t]
\setlength{\unitlength}{5mm}
\begin{picture}(12,13)(-6,-2)
\put(0,8){\vector(1,0){1}} \put(1,8){\line(1,0){1}}
\put(3,8){\circle{2}} \put(3,8){\makebox(0,0){$\vec{x}_T$}}
\put(3.8944,8.4472){\vector(2,1){1}}
\put(4.8944,8.9472){\line(2,1){1}}
\put(4,8){\vector(1,0){1}} \put(5,8){\line(1,0){1}}
\put(0.2929,5.2929){\vector(1,1){1}}
\put(1.2929,6.2929){\line(1,1){1}}
\put(3.8944,7.5528){\vector(2,-1){2.1056}}
\put(6,6.5){\line(2,-1){2}}
\put(8,5.5){\vector(1,-2){1}}
\put(9,3.5){\line(1,-2){0.5528}}
\put(8,5.5){\vector(2,1){1}} \put(9,6){\line(2,1){1}}
\put(8,5.5){\vector(1,0){1}} \put(9,5.5){\line(1,0){1}}
\put(10,1.5){\circle{2}} \put(10,1.5){\makebox(0,0){$\vec{x}_D$}}
\put(10.8944,1.9472){\vector(2,1){1}}
\put(11.8944,2.4472){\line(2,1){1}}
\put(10.8944,1.0528){\vector(2,-1){1}}
\put(11.8944,0.5528){\line(2,-1){1}}
\put(7.1056,0.0528){\vector(2,1){1}}
\put(8.1056,0.5528){\line(2,1){1}}
%
\put(0.8,8.1){\makebox(0,0)[b]{proton}}
\put(0.8,7.8){\makebox(0,0)[t]{$\vec{P}$}}
\put(6.0944,9.4472){\makebox(0,0)[l]{$p'_{T1}$}}
\put(6.2,8){\makebox(0,0)[l]{$p'_{T{n_T}}$}}
\put(5.8,8.3){\makebox(0,0)[b]{$\vdots$}}
\put(0.2929,5.1929){\makebox(0,0)[t]{T particle}}
\put(6.2,6.3){\makebox(0,0)[tr]{S particle}}
\put(7.9,5.4){\makebox(0,0)[tr]{$\vec{x}_S$}}
\put(10.2,6.5){\makebox(0,0)[l]{$p'_{S1}$}}
\put(10.2,5.5){\makebox(0,0)[l]{$p'_{S{n_S}}$}}
\put(9.9,5.6){\makebox(0,0)[b]{$\vdots$}}
\put(8.9,3.6){\makebox(0,0)[tr]{
$\stackrel{\scriptscriptstyle (-)}{\nu}_{\hskip-3pt j}(m_j)$
}}
\put(6.9056,0.0528){\makebox(0,0)[r]{D particle}}
\put(13.0944,2.9472){\makebox(0,0)[l]{$p'_{D1}$}}
\put(13.0944,0.0528){\makebox(0,0)[l]{$p'_{D{n_D}}$}}
\put(12.6,1.5){\makebox(0,0)[l]{$\vdots$}}
\end{picture}
\caption{\label{DIF} Feynman diagram for decay in flight (DIF) of
the neutrino source particle which is produced by a proton with
momentum $\vec{P}$ hitting a target (T) particle localized at
$\vec{x}_T$. In addition to the final momenta of DAR there are $n_T$
final momenta originating from the target process.}
\end{figure}
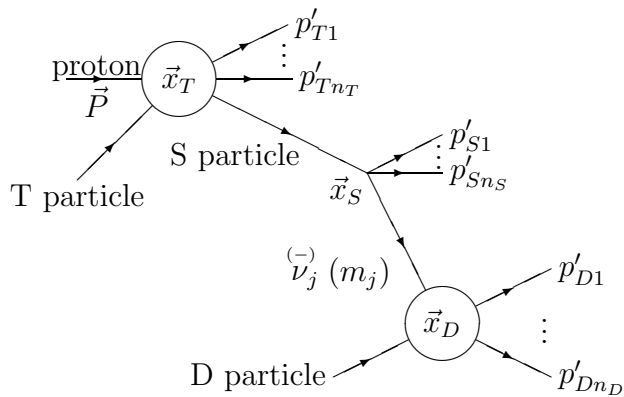

The further discussion is based on the following assumptions:
  \begin{enumerate}
  \renewcommand{\labelenumi}{\Roman{enumi}.}
  \item The wave function $\phi_D$ of the detector particle does not spread
    with time which amounts to
    \begin{equation}
    \phi(\vec{x},t) = \psi_D(\vec{x}-\vec{x}_D)\, e^{-i E_{DP}t} \,,
    \end{equation}
    where $E_{DP}$ is the sharp energy of the detector particle and
    $\psi_D(\vec{y})$ is peaked at $\vec{y}=\vec{0}$.
  \item The detector is sensitive to momenta (energies) and possibly to
    observables commuting with momenta (charges, spin).
  \item The usual prescription for the calcuation of the cross section is
    valid. 
  \end{enumerate}
With the amplitudes symbolized by Figs.~\ref{DAR} and \ref{DIF} the
oscillation probabilities are obtained by
\begin{equation}\label{Pav}
\left\langle 
P_{\stackrel{\scriptscriptstyle (-)}{\nu}_{\hskip-3pt \alpha} 
\to \stackrel{\scriptscriptstyle (-)}{\nu}_{\hskip-3pt \beta}}
\right\rangle_\mathcal{P} \propto
\int dP_S \int_\mathcal{P} \frac{d^3 p'_{D1}}{2E'_{D1}} \cdots 
\frac{d^3 p'_{D{n_D}}}{2E'_{D{n_D}}} \: \left| 
\mathcal{A}_{\stackrel{\scriptscriptstyle (-)}{\nu}_{\hskip-3pt \alpha} 
\to \stackrel{\scriptscriptstyle (-)}{\nu}_{\hskip-3pt \beta}}
\right|^ 2 \,.
\end{equation}
In this equation we have indicated the average over some region $\mathcal{P}$
in the phase space of the final particle of the detection process. If no final
particle of the neutrino production process is measured then one has to
integrate over the total phase space of these final states. By definition, at
the source (detector) a neutrino 
$\stackrel{\scriptscriptstyle (-)}{\nu}_{\hskip-3pt \alpha}$ 
($\stackrel{\scriptscriptstyle (-)}{\nu}_{\hskip-3pt \beta}$)
is produced (detected) if there is a charged lepton $\alpha^{\pm}$ 
($\beta^{\pm}$) among the final states.

In perturbation theory with respect to weak interactions, according to the
Feynman diagrams Figs.~\ref{DAR} and \ref{DIF} one has to perform integrations
$\int d^4 x_1$, $\int d^4 x_2$ and $\int d^4 q$ corresponding to the
Hamiltonian densities for neutrino production and detection and the propagators
of the mass eigenfields, respectively. 
These integrations are non-trivial because $\psi_D$
and the source (target) wave functions are not plane waves, but are localized
at $\vec{x}_D$ and $\vec{x}_S$ ($\vec{x}_T$), respectively. After having
performed the integrations over $x_{1,2}$ and $q^0$, in the asymptotic limit
$L \to \infty$ only the neutrinos on mass shell contribute to the
amplitude~\cite{GS96} which can be written as~\cite{GS96,GSM99}
\begin{equation}\label{ampinfty}
\mathcal{A}^\infty_{\nu_\alpha\to\nu_\beta} =   
\sum_j \mathcal{A}^S_j \mathcal{A}^D_j
U_{\beta j}U^*_{\alpha j} e^{i q_j L} 
\end{equation}
with
\begin{equation}\label{kin}
E_D = \sum_{b=1}^{n_D} E'_{Db} - E_{DP} \quad \mbox{and} \quad 
q_j = \sqrt{E_D^2 - m^2_j} \,.
\end{equation}
$\mathcal{A}^S_j$ and $\mathcal{A}^D_j$ denote the amplitudes for 
production and detection, respectively, of a 
neutrino with mass $m_j$. Note that $E_D$ is the energy on the neutrino line
in Figs.~\ref{DAR} and \ref{DIF} and it is independent of $m_j$. This is an
immediate consequence of the assumptions in this section. Furthermore,
due to the above-mentioned integrations and the asymptotic limit we obtain
\begin{equation}\label{AD}
\mathcal{A}^D_j \propto \widetilde{\psi}_D(-q_j \vec{\ell} + \vec{p}\,'_D)
\quad \mbox{with} \quad \vec{\ell} = (\vec{x}_D-\vec{x}_S)/L 
\quad \mbox{and} \quad \vec{p}\,'_D = \sum_{b=1}^{n_D} \vec{p}\,'_{Db} \,,
\end{equation}
where $\widetilde{\psi}_D$ is the Fourier transform of $\psi_D$.

\section{Results}

The preceding discussion leads us to the conclusion that with the assumptions
stated in Section 2 the neutrino mass eigenstates are characterized by the
energy $E_\nu \equiv E_D$ and momenta $q_j$ (\ref{kin}).
Thus they have all the same energy determined by the detection process, but the
momenta are different.
The summation over $E_D$ is incoherent, i.e., it occurs in the cross section
(see Eq.~(\ref{Pav})), not in the amplitude (\ref{ampinfty}). In this sense
there are no neutrino wave packets in experiments conforming with our
assumptions. Note that it has been
pointed out in~\cite{KNW96} that a coherent or incoherent neutrino energy
spread cannot be distinguished in neutrino oscillation experiments.
Since the neutrino energy can in principle be determined
with arbitrary precision the coherence length can theoretically be increased
solely by detector manipulations.~\cite{KNW96,GK98}
From Eq.~(\ref{AD}) it follows that with $\Delta m^2 \equiv |m^2_j-m^2_k|$
the condition
\begin{equation}\label{ACC}
|q_j-q_k| \simeq \frac{\Delta m^ 2}{2E_D} 
\lesssim \sigma_D \quad \mbox{or} \quad
\sigma_{xD} \lesssim \frac{1}{4\pi} L^\mathrm{osc}
\end{equation}
is necessary for neutrino oscillations,
where $\sigma_D$ and $\sigma_{xD}$ are the widths of the wave function of the
detector particle in momentum and coordinate space, respectively, and 
$L^\mathrm{osc}$ is the oscillation length.~\cite{kayser,rich,GS96} In
realistic experiments condition (\ref{ACC}) holds because $\sigma_{xD}$ 
is a microscopic whereas $L^\mathrm{osc}$ a
macroscopic quantity. For DAR an analoguous condition exists for the width of
the neutrino source wave function. 

For more details of the field-theoretical approach to neutrino oscillations,
for a consideration of the finite lifetime of the neutrino source in the case
of DAR and for an application of the results to the LSND and KARMEN 
experiments we refer the reader to~\cite{GS96,GSM99}. We have shown that in
these experiments effects of the finite liftetime can be neglected.
For a discussion of DIF see~\cite{campagne}. Thus in the framework discussed
here all corrections to
Eq.~(\ref{P}) are negligible. Note that we have not taken into account or
discussed the
interaction of the final state particles in the source with the environment
(``interruption of neutrino emission''), a possible intermediate range of the
asymptotic limit $L \to \infty$ as found in~\cite{ioan} and the possibility
that in some cases (e.g., the KARMEN experiment) it is not realistic to 
use the conventional procedure to calculate the cross section (\ref{Pav}) by
taking the asymptotic limit of the final time to infinity.

\end{document}